\documentclass[16pt]{article}
\usepackage{amsmath}
\usepackage{amssymb}
\usepackage{graphicx}
\usepackage{color}
\usepackage{verbatim}
\usepackage{float}
\usepackage[small]{caption}
\usepackage{subfig}
\usepackage[a4paper, left=2.5cm, right=2.5cm, top=3.5cm, bottom=3.5cm, headsep=1.2cm]{geometry}

\def\Log{{\rm{Log}\,}}

\newcommand{\be}{\begin{equation}}
\newcommand{\ee}{\end{equation}}
\newcommand{\bea}{\begin{eqnarray}}
\newcommand{\eea}{\end{eqnarray}}
\newcommand{\bal}{\begin{align}}

\newcommand{\enl}{\end{align}}

\def\ov{\over  }

\begin{document}


{\bf{ A Theory-independent Way of Unambiguous Detection of  Wino-like particles at LHC}}
\vskip0.5cm

G.Calucci, Dept. of Physics, Trieste University, R.Iengo, SISSA Trieste

\vskip0.5cm

{\bf{Abstract. }} We propose to use the change of the energy lost by ionization, measured by silicon detectors,  before and after  the passage through a bulk of dense matter,  for unambiguously detecting  highly massive single-charged particles, which could be produced at LHC, in particular Winos with mass in the TeV range, whose c-tau is expected to be some cms long, although the method is also efficient for masses down to 10GeV. For convenience, a QED derivation of the modern version of the Bethe-Block formula is also provided.

\vskip0.5cm

Here we describe a proposal of a device for unambiguously detecting sufficiently long-lived ($c \tau\sim$ some centimeters)
highly massive single-charged particles, which could be produced at LHC. In particular, we have in mind the possible detection of Winos, having 
charge 1 and 
$c \tau \sim 7-14 cm$, depending on the mass, from a 2-loop computation reported in ref(\cite{jap}), which decay into nearly in-mass-degenerate neutralinos that are candidates for dark matter. But of course the method would be suitable for detecting any new massive charged particle with similar or greater lifetime.

In the literature, there are descriptions of devices for detecting by ionization highly ionizing particles (say $charge/\beta\geq 5$ ref.(\cite{MoEDAL})
and/or very long-lived particles ($c\tau\geq$ meters), see ref(\cite{atlaslongl} ), therefore with a scope rather different from ours.

The idea is to use the ionization produced by the passage in matter and measured by silicon detectors
for discriminating between the lighter and the heavier particles which could be produced by the beam interactions at LHC 
with similar initial $\beta$, and therefore for suppressing the background.

In fact, while the energy lost by ionization ${dE\ov dx}$ mainly depends on $\beta$, the same amount of momentum loss in $dx$ 
will reduce the lighter particles' $\beta$ more than heaviers' one, therefore inducing an even larger successive energy loss and making 
$\beta$ smaller and smaller for the lighter particles.

By inserting between successive silicon detectors a thick layer of a dense substance (gold, in our example),
one forces the lighter particles to reduce their velocity more than the heavier ones.

As a result, the ionization measured in the next silicon detector will be significantly larger for the lighter particles with respect to the heavier ones.

Moreover, a particle with a high mass will ionize in nearly same way all the silicon detectors.  Therefore, it will provide a quite unambiguous signal, very different from the background due to the passage of known particles.

We will see that the results of the modern version of the Bethe-Block formula indicate that it would be possible to discriminate new particles from the known ones for masses $\geq 10 GeV$.

This device could also give an information on the value of the mass of a new particle or provide a lower limit on it.

In fact we will see that, by making use of  the modern version of the Bethe-Block formula,
for a mass = 100GeV the difference of ionization between the last and the first silicon detectors is a $\sim 5\%$ of the total,
that could be near to the confidence limits of the measure, whereas for higher masses the ionization in  
the successive silicon detectors is almost the same and moreover mass independent.

Therefore, this device could give the mass value of the particle from the measure of the ionization in the silicon detectors
for masses up to $\sim$ 100GeV, otherwise providing just a lower limit.

\vskip0.2cm
In conclusion, the unambiguous signal of a highly massive  charged particle is an almost constant ionization in the successive silicon detectors. Every other particle will produce a quite significant increase of the ionization.

In particular, a 3TeV Wino, that is the  nearly degenerate charged partner of the neutralino in the mass-range compatible with the observed dark-matter abundance \cite{ur}, will give a constant ionization in all the silicon detectors.

\vskip0.2cm

Actually, the innermost silicon detector taking place in the present LHC devices only signals the passage of a charged particle, providing important informations, for instance on the trajectory, but it does not measure the deposited ionization\footnote{We thank Dr. Susanne Kuehn for a correspondence on this point.}.

Therefore, what we present here is an idea which could be possibly implemented when seen technically viable by the instrumentation experts.

Anyhow, we think it is useful to discuss our ideal proposal, to see in some detail what efficacy it could have in the detection. It could  suggest some alternatives to the more standard ways of revealing new particles.

\vskip0.2cm

We describe here an example of what we have in mind. That is, a cylinder coaxial with the beam around the collision region, made of 5 cylindrical strata of different material and thickness, and for different purposes.

First, a a cylindrical silicon detector of 0.1cm thickness, then a  0.4cm  thick cylindrical layer of gold, then another 0.1cm thick silicon detector, 
then a 0.3cm thick cylindrical layer of gold, and a final 0.1cm thick silicon detector.
Only the silicon is meant to detect ionization.

Of course, one could also imagine a simpler device, as just two silicon detectors, with say a 0.3cm thick layer of gold in between, would suffice in principle, and it would maybe easier to fit in. We nonetheless go on with the previous, more disambiguating, recipe.

In Fig.1 we show a transverse section of the device, with the silicon coloured in light blue and the gold in orange.

The total thickness of the device being 1cm,  this device could detect long-lived particles having $c \tau \sim 7 cm$ (as expected for a Wino with a mass larger than 250 GeV, \cite{jap}) giving $ c \tau \beta/\sqrt{1-\beta^2}=3 cm$ for $\beta=0.4$ and $ c \tau \beta/\sqrt{1-\beta^2}=5.4 cm$ for $\beta=0.6$,
if put at a distance of $\sim 3cm$ from the beam interaction, 
which we understand to be so far the minimal distance from the beam of a detector see ref(\cite{derosa}).
Particles having $c\tau=14 cm $ (as expected for a Wino with a mass around 100GeV, \cite{jap})  give
$ c \tau \beta/\sqrt{1-\beta^2}=6.2 cm$ already for   $\beta \geq\sim 0.4$.

\begin{figure}[H]
{\label{Fig.1}
		 \includegraphics[width= 3 in]{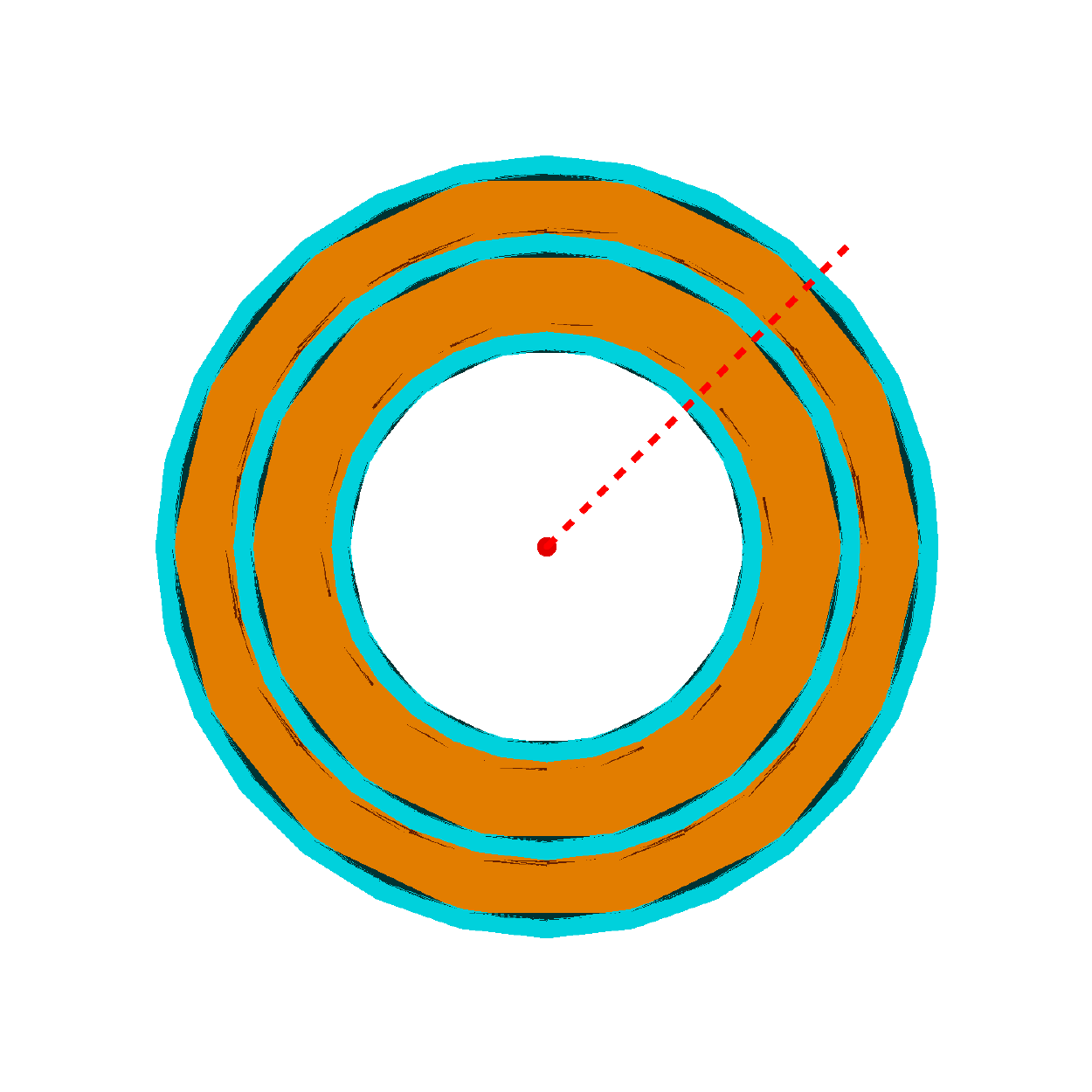} }
		 \caption{transverse section of the detecting device. Light-blue:silicon, orange:gold}
		 \vskip0.4cm
		\end{figure}

By using the so-called Bethe-Block formula  (in the form which can be found in modern references, that is eq.(\ref{befin}) of the Appendix) we solve numerically the differential equation for the momentum of a particle as it goes trough the device and compute the ionization deposited in the silicon detectors.

Since the difference between the ionization energy of a Si atom and the energy required to move a carrier to the conduction band is very small as compared to the particle's energy, 
moreover the possible doping concentration being much less than the Si atoms one, we have taken the average Si atom ionization
as the main source of the energy deposited, see Appendix. Anyhow,  including additional ionization would enhance the effect.

In the Appendix we present, for convenience of the reader, a pedagogical derivation  from QED  of the modern version of the Bethe-Block equation.

\vskip0.2cm

We have done the exercise for a heavy particle of mass $= ~3TeV,~100GeV,~50GeV,~10GeV$ and for the long-lived barion $\Xi^-$, which we take as a benchmark for the possible background since it
is the longest lived 
($c\tau=4.91cm$ ) known particle of high mass.

Any other known sufficiently long-lived charged particle (say $\Sigma$, proton, pion, lepton) is lighter and therefore its ionization pattern  will make the discrimination more evident ( $\Omega^-$ is heavier than $\Xi^-$ but it has a low $c\tau=2.46 cm$).

Also, the computation has been made for purely transverse particle trajectories. Tilted trajectories would increase the amount of crossed matter and therefore would increase the effect.

\vskip0.2cm

In Fig.2 we show the ionization in the successive silicon detectors of  ionizing particles of mass $= ~ 10 GeV,~ 50 GeV, ~ 100 GeV,~ 3 TeV$, for initial $\beta=0.4$ and $0.6$.

\begin{figure}[H]
                {\label{Fig.2}
		 \includegraphics[width= 3 in]{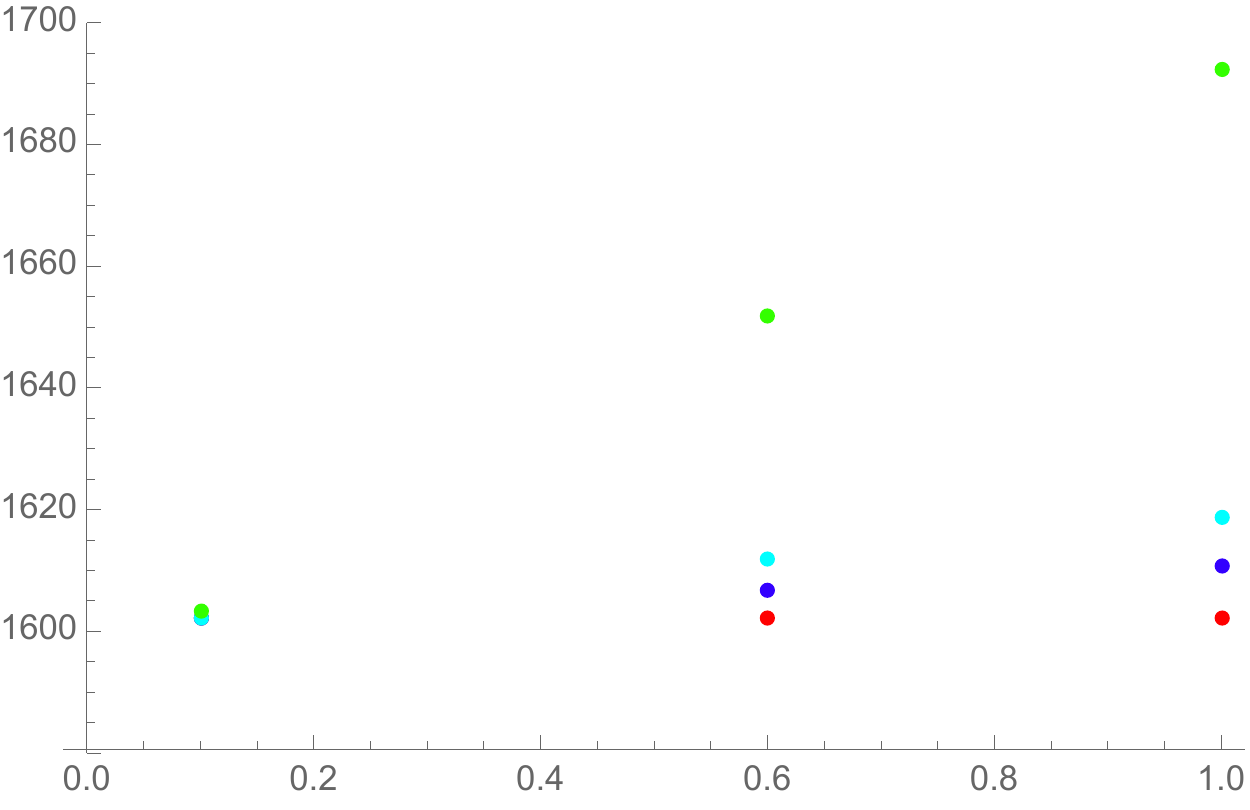} 
		 \includegraphics[width= 3 in]{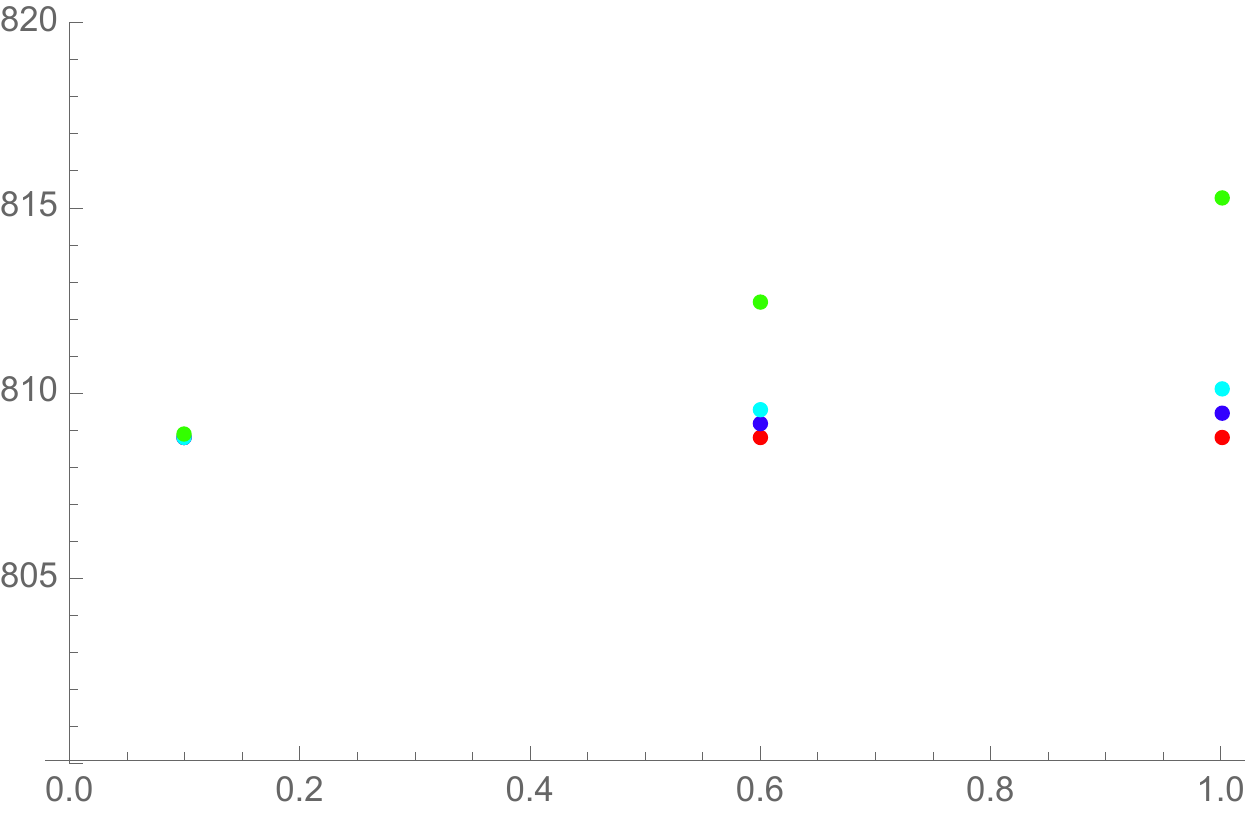} }
		 \caption{ionization in successive silicon detectors  for particles mass  10 GeV, 50 GeV,  100 GeV,  3 TeV, 
		 for initial $\beta_{in}=0.4$ (left) 
		 and $0.6$    (right, note the different scale in the y-axis). The larger the mass the more constant ionization.               
		 y-axis: ionization  in KeV, x-axis: Si detector position from the inner layer in cm. }
		 \end{figure}

The main point is that a $3TeV$ ionizing particle (call it Wino for short) loses a so small amount of energy, with respect to the incident momentum, that it maintains almost exactly its incoming $\beta_{in}$ and therefore its ionization signal in the successive silicon detectors is quite the same.
This also essentially  happens for massive particles down to $10 GeV$, in which case the difference in ionization is not so small but still much less than what occurs for the known particles.

All the other known long-lived particles lose so much energy that their $\beta$ significantly decreases and the ionization significantly increases.
For low $\beta$ or for low mass, these particles do not even reach the next silicon detector, as they stop in the gold before.
This we have found to happen  for $\Xi^-$ at $\beta\leq 0.3$.

Of course, this effect diminishes for very high $\beta$, because both the high-mass particle and the $\Xi^-$ ionizations in the successive silicon detectors decrease and approach each other.
But this would be the region in which the high-mass particle production will be more rare, and therefore one could disregard those cases by putting an upper cut on the  ionization to be recorded.
Note that the difference in the (transverse) momenta between the high-mass Wino-like and the known particles grows with $\beta$. Therefore, the observation of 
an anomalous high momentum could complement the ionization measures for high $\beta$.

\vskip0.2cm

In Table.1 we report  in numbers the sample of our results shown in Fig.2 together with the results for our benchmark $\Xi^-$. For completeness, we also show our results for $\Omega^-$ although probably not relevant due to the short $c\tau$. The ionization is in KeV. 

We have  estimated the statistical uncertainty of the ionization computation, by computing by eq.(\ref{varian}) for each particle and each $\beta$ the variance of the energy, that is the average $~<(p_0 ~- <p_0>)^2>~$ as a function of $x$ (the distance crossed in the matter),  and then recomputing the ionization in the detectors for 
$<p_0> \pm \sqrt{<(p_0 ~- <p_0>)^2>}$. 
We find that this uncertainty is quite small for the high-mass particles and it can barely seen in the previous Fig.2. 
In Table 1 we indicate the uncertainty in the ionization only for the $\Xi^-$ and $\Omega^-$ particles. 

\begin{center}
\begin{tabular}{| c || c  c  c | | c  c c |}
\hline
\multicolumn{7}{| c |}{Table 1. Energy lost (KeV) in the Si detectors} \\
\hline
  &  & $\beta_{in}=0.4$ &  &   & $\beta_{in}=0.6$  &  \\
 \hline   
   mass       & $1^{rst} Si$ &$2^{nd} Si$ & $3^{rd} Si$ &$1^{rst} Si$ &$2^{nd} Si$ &$3^{rd} Si$ \\
\hline
\hline
$ 3000 GeV $ & 1602 & 1602 & 1602 & 809 & 809 & 809 \\
\hline
$ 100 GeV$ & 1602 & 1607 & 1610 & 809  &  809 & 809 \\
\hline
$ 50 GeV$ & 1602 & 1611 & 1619 & 809 & 809 & 810 \\
\hline
$ 10 GeV$ &  1603 & 1652 & 1692 & 809 & 812 & 815 \\
\hline
$ \Xi^- ~1.321 GeV $ & 1610 $\pm$  2 & 2205 $\pm$ 32 & 4596 $\pm$ 231 & 809 $\pm$ 0.2 & 838 $\pm$ 3 & 864 $\pm$ 4 \\
\hline
$\Omega^- ~ 1.672 GeV $ & 1608 $\pm$ 1 & 2015 $\pm$ 20 & 2787 $\pm$ 58 & 809 $\pm$ 0.2 & 832 $\pm$ 2 & 851 $\pm$ 3 \\
\hline
\end{tabular}
\end{center}

\vskip0.2cm

By taking the $\Xi^-$ as a benchmark,
in Fig.3 we show our results for the ionization difference in each of the three silicon detectors
\be
\Delta(ionization)=(ionization[\Xi^-] - ionization[highmass])     \nonumber
\ee
for  $\beta=0.4,0.45.0.5,0.55,0.6$ and when the high-mass particle is 3 TeV massive (our best choice for a Wino) and when it is 10 GeV massive.
We see that the results for the two cases are very similar. 

We show two points for each result, by adding and subtracting the statistical uncertainty, computed as above said, to the ionization difference.

In the first detector the ionization difference is small 
and it does not depend much on $\beta$, whereas it significantly increases in the second and third detector. This difference diminishes for increasing $\beta$, as expected, because for larger $\beta$ there is less ionization loss.

But still at $\beta=0.6$ the ionization difference is an amount of tens of KeV which could unambiguously detected.

\begin{figure}[H]
                     {\label{Fig.3}
		 \includegraphics[width= 3 in]{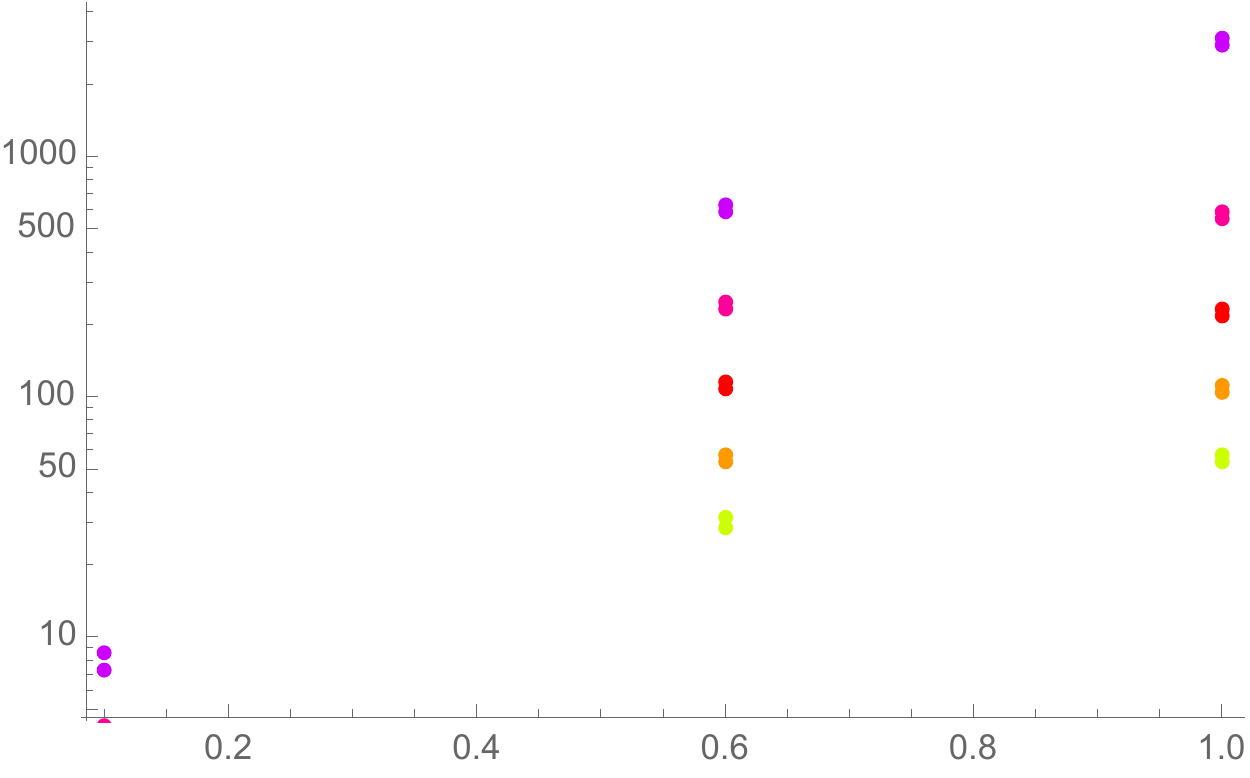} 
		 \includegraphics[width= 3 in]{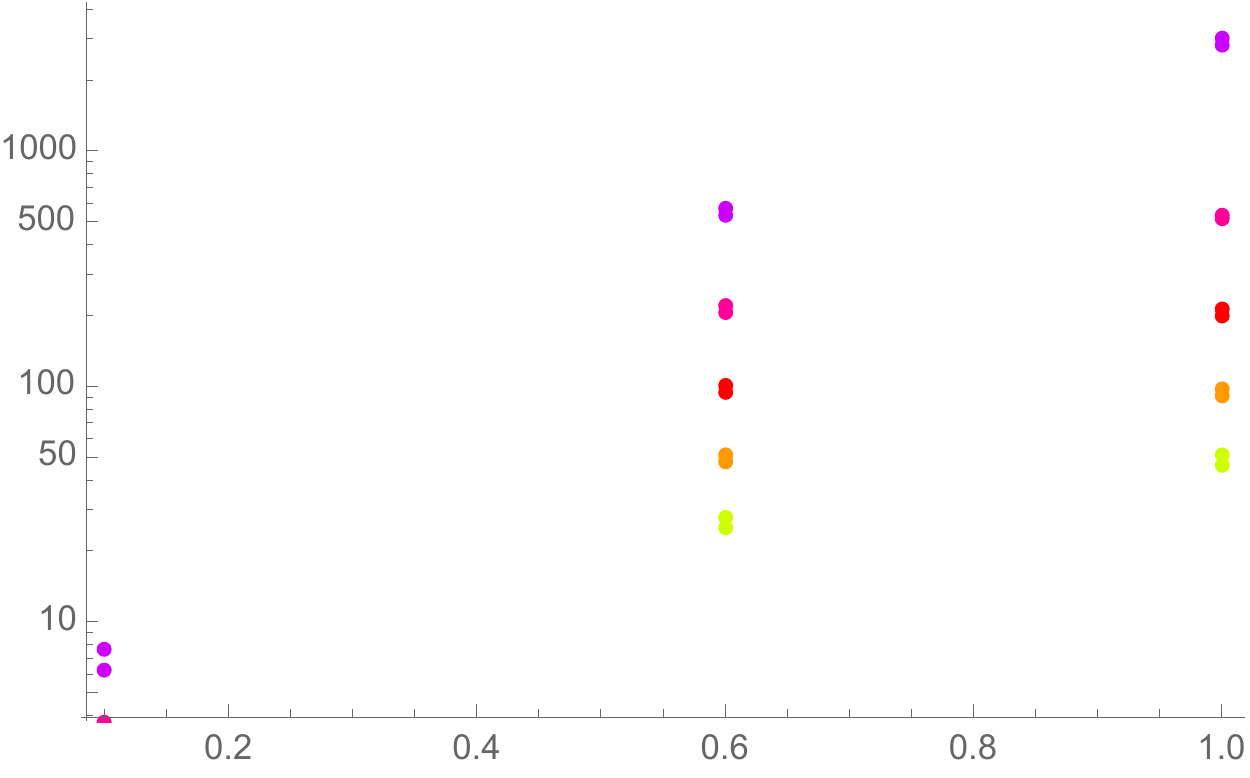} }
		 \caption{ionization[$\Xi^-$] - ionization[highmass] in the succesive silicon detectors,  for initial $\beta_{in}=0.4,0.45.0.5,0.55,0.6$ (smaller $\beta$ larger ionization difference)
		 for highmass=3 TeV (left), and for highmass=10 GeV (right). Double/overlapping points indicate the estimated uncertainty. 
		 y-axis: ionization difference in KeV (logarithmic scale), x-axis: Si detector end position from the inner layer in cm. }
		 \end{figure}

\vskip0.2cm

Finally, in the literature there is some debate whether, for thin silicon detectors, the fluctuations are gaussian distributed (as we implicitly assumed in estimating the uncertainties) or 
there are long tails and it would be better to use the formula eq.(\ref{most}) of the Appendix for the most probable energy loss $\Delta E_{most}$ in a width $\Delta x$.

 Actually, U.Fano \cite{fano} points out that the probability distribution is expected to be gaussian whenever the energy lost in the interval $\Delta x$ is much larger than the maximum loss in a single collision.
We have seen that this is be true in our case, validating our results. 

But just for testing the robustness of our results, we have repeated the computation by assuming for the silicon detectors the formula eq.(\ref{most})  for 
$\Delta E_{most}$. While in general we get in this way 
a ionization in silicon to be some $90 KeV$ less than the Bethe-Block result (except when it is very large), 
the ionization difference $\Delta_{ionization}$ remains essentially the same.

As an illustration, in Fig.(4) we compare (ionization[$\Xi^-$] - ionization[10GeV]) as computed by the Bethe-Block formula eq(\ref{befin}) of the Appendix with
$(\Delta E_{most}[\Xi^-] - \Delta E_{most}[10GeV])$ as computed from eq.(\ref{most}).

\begin{figure}[H]
                    {\label{Fig.4}
		 \includegraphics[width= 3 in]{resfig10g.pdf} 
		  \includegraphics[width= 3 in]{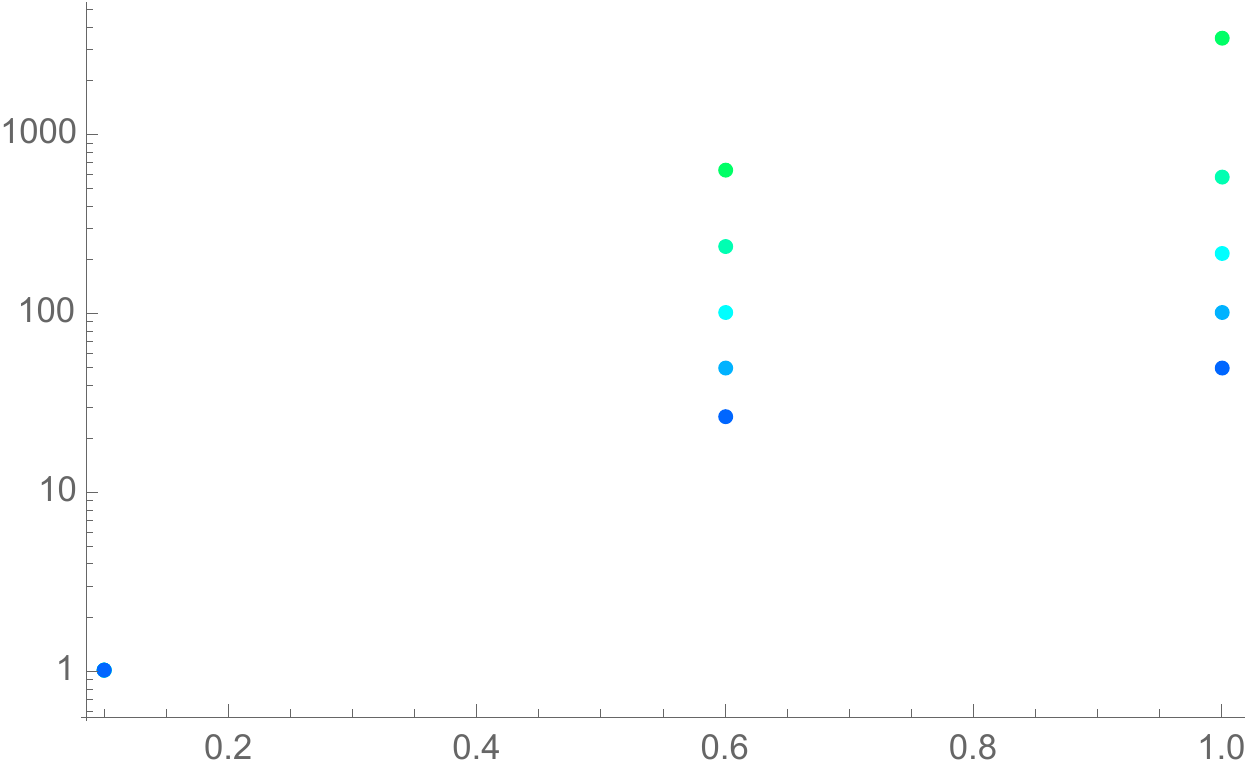} }
		 \caption{ionization[$\Xi^-$] - ionization[10GeV]  in the succesive silicon detectors,  for initial $\beta_{in}=0.4,0.45.0.5,0.55,0.6$,
 computed from Bethe-Block (left) and from the most probable ionization  eq(\ref{most}) (right).
		 y-axis: ionization difference in KeV, x-axis: detector position from the inner layer in cm. }
		 \end{figure}

In conclusion we find that in any case an event where the ionization is sufficiently high to indicate a $\beta\leq\sim 0.6$, 
but the difference in the ionization before and after the passage through a bulk of dense matter is
anomalously small, would clearly 
indicate the occurrence of a new high-mass particle.


\vskip0.5cm

\noindent{\bf{Appendix. \\
A Pedagogical Derivation of the Bethe-Bloch Formula (modern version) from QED.}}

\vskip0.3cm
We begin by following  U.Fano ref(\cite{fano}).

Introduce the distribution $F(E,x)$ of the incoming ionizing particles (call them "Wino")  having energy $E$ at a distance $x$ from the beginning of the material. \\
$\int{dE F(E,x)}={\cal{N}}_w$ is the total number of Winos and it is independent of $x$, neglecting here the possible Wino decay.

The number of collisions of the incoming ionizing particle (of energy $E$) per unit length and  per unit energy loss during its passage through matter will be
\be
 {d^2N(E,\Delta)\ov dx d\Delta}={d\sigma(E,\Delta)\ov d\Delta} \rho_N    
\ee
where $\sigma$ is the ionizing cross-section and $\Delta =E-E'$ is the difference  the energy before ($E$) and after ($E'$) the collision. Notice that ${d\sigma\ov d\Delta}$ in general depends on $E$ and $\Delta$.
 $\rho_N={dn\ov dV}$ is the number density of the (bound) electrons. If the atomic number is $Z$ and the atomic weight is $A$, the number density of the electrons is 
 $\rho_N={ \rho Z \ov A m_{pr}}$ where $\rho$ is the mass density of the material and $m_{pr}$ is the proton mass (neglecting the neutron-proton mass difference).

$F(E,x)$ will depend on $x$ because of the interaction with the electrons of the material:
\be
F(E,x+dx)= F(E,x)+dx \Big\{ \int_0^\infty d\Delta {d^2N(E+\Delta,\Delta)\ov dx d\Delta } F(E+\Delta,x)-\int_0^{E-m_w} d\Delta {d^2N(E,\Delta)\ov dx d\Delta } F(E,x) \Big\}
\ee
that is, after the interval $dx$ , an amount $dx \int{  d\Delta {d^2N(E,\Delta)\ov dx d\Delta } F(E,x) }$ of particles  with energy $E$  disappears due to the interaction with the material (their energy after the interaction being  $E-\Delta$; the upper limit for $\Delta$ insures that $E\ge m_w$)
and are replaced by an amount  $dx \int{  d\Delta ({d^2N(E+\Delta,\Delta)\ov dx d\Delta } F(E+\Delta,x)}$ of particles that loose  the energy $\Delta$ from their initial energy $E+\Delta$ .

Note that by integrating over $E$ 
\bea
\int_{m_w}^\infty dE \int_0^\infty d\Delta {d^2N(E+\Delta,\Delta)\ov dx d\Delta } F(E+\Delta,x) = \int_0^\infty d\Delta\int_{m_w+\Delta}^\infty dE' {d^2N(E',\Delta)\ov dx d\Delta } F(E',x) \\ \nonumber
\int_{m_w}^\infty dE\int_0^{E-m_w} d\Delta {d^2N(E,\Delta)\ov dx d\Delta } F(E,x) = \int_0^\infty d\Delta\int_{m_w+\Delta}^\infty dE {d^2N(E,\Delta)\ov dx d\Delta } F(E,x) \nonumber
\eea
hence $\int_{m_w}^\infty dE F(E,x+dx)= \int_{m_w}^\infty dE F(E,x)={\cal{N}}_w$.

\vskip0.3cm

Therefore
\be
{d {\cal{P}}(E,x)\ov dx}= -{\cal{P}}(E,x) \int{d\Delta {d\sigma(E,\Delta)\ov d\Delta}}\rho_n+\int{ d\Delta {\cal{P}}(E+\Delta,x) {d\sigma(E+\Delta,\Delta)\ov d\Delta } \rho_n} \label{C}
\ee

where we have introduced the normalized probability distribution 
${\cal{P}}(E,x)={F(E,x)\ov {\cal{N}}_w}$.

By multiplying both sizes by $E$ and integrating over $E$ we get:
\be
{d<E>(x)\ov dx}=-<\int{d\Delta \Delta {d\sigma(E,\Delta)\ov d\Delta } \rho_N}>
\ee

By approximating $<{\cal F}(E)>\sim {\cal F}(<E>)$, for the appropriate function  ${\cal F}(E)$, we get ($E_{av}=<E>$)
\be
{dE_{av}(x)\ov dx}=-\int{d\Delta \Delta {d\sigma(E_{av}(x),\Delta)\ov d\Delta } \rho_N} \label{D} 
\ee
This is the basic formula for the energy lost by ionization of a particle passing through matter.

In the following, we will  write $E$ in the place of $E_{av}$ in eq.(\ref{D}), as it is usually done  forgetting fluctuations.
 
We can evaluate the fluctuations 
by further multiplying both sides of eq(\ref{C}) by $E^2$ and integrate over $E$ to get
\be
{d<E^2>\ov dx}=-2<E\int{d\Delta \Delta{d\sigma(E,\Delta)\ov d\Delta } \rho_N}>+<\int{d\Delta \Delta^2 ~{d\sigma(E,\Delta)\ov d\Delta } \rho_N}>
\ee

By the previous approximation we get
\be
{d~ var[E](x)\ov dx}=\int{d\Delta \Delta^2~{d\sigma(E_{av}(x),\Delta)\ov d\Delta } \rho_N} \label{E}
\ee
where $var[E]\equiv <(E-<E>)^2>=<E^2>-<E>^2$.

One can solve the differential equation eq.(\ref{D}) for $E_{av}(x)$ and use the solution for computing $var [E](x)$ by integrating over $x$ the r.h.s of eq.(\ref{E}).

Our aim is to compute the r.h.s. of eq.(\ref{D}), and then also the r.h.s. of eq.(\ref{E}).

\vskip1cm

In the LAB, we consider the incoming (ionizing) particle  (eventually this particle could be taken to be {\cal{Wino}}) with mass $m_w$ and momentum $p_{\mu}=\{p_{0},\vec p\}$ colliding with a (bound) electron of mass $m_e$.
After the interaction the Wino 3-momentum will be ${\vec p}~'$ and the 3-momentum transfer is
\be
\vec q=\vec p-{\vec p}~'
\ee

The binding energy of the electron is very small (of the order of some ten eV) compared to  the other energy scales, say the masses $m_e,m_w$ and the incoming momentum $p_w$. 
The minimum value of $\Delta$ in the integration appearing in the r.h.s. of eq.(\ref{D}) will be of the order of the binding energy and its precise value and also the precise 
expression of the integrand in this low $\Delta$ region could depend on the actual dynamics of the electron-atom system.

 On the other hand, outside that region one can neglect the binding altogether and consider a free electron initially at rest and take the standard expression of the relativistic Rutherford cross-section.
 
 Therefore, we will first discuss the low $\Delta$ region, where $q$ is also small,  working in the Lab frame. In this region the dynamics of the electron is non-relativistic, while the incident particle has to be treated in general as relativistic.
 
 The cross-section for exciting the electron from the ground state to a level $n$  (with energy $E_n$, by convention ground state energy $=0$) is
 \bea
 d\sigma_n &=& {1\ov \beta} |<{\vec p}~' n|V|\vec p~ 0>|^2 {d{\vec p}~' \ov (2\pi)^6}  (2\pi)^4 \delta (p'_{0}+E_n-p_{0}) \\ \nonumber
 &=& {1\ov (2\pi)^2\beta^2} |<{\vec p}~' n|V|\vec p~ 0>|^2 {p~'}^2   ~d(\cos\theta) d\phi
 \eea
 where $\beta=p/p_{w0}\sim \beta'=p~'/p'_{w0}$ since we are considering the low $q$ region.
 
 Now from  $p_0=p'_0+E_n$ we get  $p^2={p~'}^2+2 p'_0 E_n+O(E_n^2)$ That is at the leading order in $E_n$
 \be
 p-p~'= {p_0\ov p}E_n , ~~ q^2 =({E_n\ov\beta})^2+2 p p~'(1-\cos\theta)
 \ee
 Therefore
 \be
 q^2_{min}=({E_n\ov\beta})^2
 \ee

 It is convenient here to write the Rutherford cross-section in the Coulomb gauge 

 \be
 <{\vec p}~' n|V|\vec p~ 0>=e^2 Z^{1/2} \big( {F_n(q)\ov q^2}+{\vec\beta_t\cdot \vec G_n(q)\ov q^2-E_n^2} \big) \label{Cou}
 \ee
where by definition  $\vec\beta_t\cdot\vec q=0$, and for small $q$
\bea
F_n(q)=Z^{-1/2} <n|\sum_j e^{i\vec q\vec r_j}|0> \sim Z^{-1/2} i q <n|\sum_i x_j|0> ~~~~~~~~~~~~~~~~~~~~~~~~~~~~~~~~~~~~~~~~~~~~~~~~ \label{F} \\
\vec\beta_t\cdot\vec G_n(q)=Z^{-1/2}\vec\beta_t\cdot<n|\sum_j \vec\alpha_j e^{i\vec q\vec r_j}|0> \sim\beta_t Z^{-1/2}i<n|\sum_j v_{jy} |0>=
\beta_t E_n Z^{-1/2}i <n|\sum_j y_j |0> \label{G}
\eea
where the $r_j,v_j$ are the position and the velocity of the $j$-electron, $x_j,y_j$ the component of the position along $\vec q$ and $\vec\beta_t$ respectively.
 We have neglected higher order terms $O(v_j q_{j'})$ in the computation of $F_n$ and $G_n$.
Note that eq.(\ref{Cou}) treats  fully relativistically the ionizing particle: the incident spinor current appearing in the  Feynman graph in the small angle approximation, $\vec\beta_{in}\sim\vec\beta_{out}$, 
is $\{{1\ov\sqrt{1-\beta^2}},{\vec\beta\ov\sqrt{1-\beta^2}}\}$, the overall factor $\gamma={1\ov\sqrt{1-\beta^2}}$ being absorbed by the normalization of the initial and final Wino.

When taking the square modulus of the amplitude and summing over $|n>$ the interference term vanishes because ($v_t$ and $\vec q\cdot\vec r$ comute)
\be
\sum_n <0| e^{-i\vec q\vec r}v_t|n><n|e^{i\vec q\vec r}|0>=<0|v_t|0>=0
\ee
by averaging over any possible direction characterizing $|0>$. 

Therefore we get
\be
d\sigma_n={ \pi e^4 Z\ov (2\pi)^2\beta^2} \big({|F_n(q)|^2\ov q^4}+{|\vec\beta_t\cdot \vec G_n(q)|^2\ov (q^2-E_n^2)^2} \big) d(q^2)
\ee
using $\int d\phi=2\pi$, ${p'}^2 d(\cos\theta)\sim p p' d(\cos\theta)=d(q^2)/2$.

Further, from eqs.(\ref{F},\ref{G})
\be
|F_n(q)|^2={q^2\ov 2 m_e}{f_n\ov E_n} ~~~~ |\vec\beta_t\cdot\vec G_n(q)|^2={\beta_t^2\ov 2m_e} E_n f_n
\ee
where $f_n=2 m_e E_n Z^{-1}|<n|\sum_j x_j|0>|^2 =2 m_e E_n Z^{-1}|<n|\sum_j y_j|0>|^2 $.

Putting by convention $H|0>=0$ one derives  the sum rule
\be
 \sum_n f_n =\sum_n 2 m_e E_n Z^{-1}<0|\sum_j x_j|n> <n| \sum_l x_l |0>=2 m_eZ^{-1}<0|\sum_j  x_j H \sum_l  x_l |0>=1
\label{SR}
\ee
 
Now, remembering that $p' =p-{E_n\ov \beta}$ so that for $\theta$ small  $p-{\vec p\ov p}\cdot{\vec p}~'={E_n\ov \beta}$,
we can write
\be
\beta_t^2 = \beta^2-{(\vec\beta\cdot \vec q)^2\ov q^2})=\beta^2-{\beta^2 (p-{\vec p\ov p}\cdot{\vec p}~')^2\ov q^2}) 
= \beta^2(1-{q^2_{min}\ov q^2})
\ee
 We get in conclusion for the total energy lost in this low $q$ region
 \be
\sum_n E_n d\sigma_n={2 \pi \alpha^2 Z\ov m_e  \beta^2}\sum_n f_n\big({1\ov  q^2}+{\beta^2(1-{q^2_{min}\ov q^2})E^2_n \ov (q^2-E_n^2)^2} \big) d(q^2)
\ee
In the second term it is convenient to introduce the variable $\cos^2\psi=q^2_{min}/q^2$ so that $dq^2=-q^4\beta^2/ E_n^2 d\cos^2\psi$.

In the first term we integrate over $q^2$ from $q^2_{min}=({E_n\ov \beta})^2$ up to some intermediate $q^2_1$ and in the second term we integrate over
$\cos^2\psi$ from ${E_n^2\ov \beta^2 q^2_1}$ to $1$.

In the limit $E_n^2<<q^2_1$ we get
\be
 \int \sum_n E_n {d\sigma_n} ={2 \pi \alpha^2 Z\ov m_e  \beta^2}\sum_n f_n\big(\log({\beta^2 \gamma^2 q^2_1\ov E_n^2})-\beta^2 \big)
 \ee
 
 By taking the set $\{f_n\}$ as a probability distribution, since  $f_n>0$ and $\sum f_n=1$, one defines the average ionization energy $I$ by 
 \be
 \log I=\sum_n f_n \log E_n
 \ee
   and finally gets the energy lost in this region\footnote{An essentially similar derivation of the first term in the r.h.s. of eq(\ref{lost1}) is found in the classic book Landau-Lifschitz Quanum Mechanics  \cite{LLB}  chapter XVIII sect.149.}

 \be
\sum_n \int_{E_n^2\ov \beta^2}^{q_1^2} dq^2 E_n {d\sigma_n\ov dq^2}={2 \pi \alpha^2 Z\ov m_e  \beta^2} \big(\log({\beta^2 \gamma^2 q^2_1\ov I^2})-\beta^2 \big)\rho_N \label{lost1}
 \ee
 
 \vskip1cm
 
Next, we leave ref(\cite{fano}) and we go on with the fully relativistic QED. 

Assuming that $q_1$ is much higher than the ionization energy $I$, in the region $q>q_1$  one can neglect  the binding of the electron and treat it as free and initially at rest.
 
 In this case we use the fully relativistic formula for the Rutherford scattering 
 (see Landau-Lifshitz: Relativistic Quantum Mechanics \cite{LLA}, problem 6 eq.(4) in chapt. IX, par.82 )
 \be
{d\sigma\ov dt}= {\pi\alpha^2\ov  m_e^2  p^2} {1\ov t^2}(4 m_e^2 p_{0}^2+ t s+{t^2\ov 2})
\label{sigma}
\ee
 where $t$ is the square momentum transfer 
 \be
 t=(\sqrt{m_e^2+q^2}-m_e)^2-q^2=-2 m_e (\sqrt{m_e^2+q^2}-m_e)
 \ee
 Note that $\Delta$, the energy lost by the ionizing particle absorbed by the electron, is 
 \be
 \Delta\equiv \sqrt{m_w^2+p^2}-\sqrt{m_w^2+(p-q)^2}={|t|\ov 2 m_e}
 \ee
 
 Therefore the energy lost in $d|t|$ in this region is
 \be
 {|t|\ov 2 m_e} {d\sigma\ov dt}\rho_N = {1\ov |t|} {\pi\alpha^2\ov 2  m_e^3 p^2}(4 m_e^2 p_{0}^2-s|t| +{|t|^2\ov 2})\rho_N
 \label{lostr}
 \ee
 
 For $q_1$ much less than $m_e$, the minimal value of $|t|$ in this region is
 \be
 |t|_1=  2 m_e (\sqrt{m_e^2+q_1^2}-m_e)  \sim q_1^2
 \ee
 As for $|t|_{max}$, it is more easily computed  in the CM frame. Neglecting $I$, the invariant square momentum transfer $t$ is in the CM:
\be
t=-2 q^2_{CM}(1-\cos(\theta)
\ee
where $q^2_{CM}$ is the solution of 
\be
s\equiv m_w^2 +m_e^2+2 m_e p_{0} =(\sqrt{q^2_{CM}+m_e^2}+\sqrt{q^2_{CM}+m_w^2})^2
\ee
giving
\be
q^2_{CM}=m_e^2{p_{0}^2-m_w^2\ov s}=m_e^2{{ p}^2\ov s}
\ee
The  maximum for $|t|$ is obtained for $\theta=\pi$ therefore\footnote{We rename $Q_{max}$ the maximum energy transfer in a single collision to match the notation of \cite{Striganov}, the same quantity being called $W_{max}$ in \cite{passage}.}
\be
|t|_{max}=4m_e^2{{ p}^2\ov s}~ \to Q_{max}\equiv  \Delta_{max}={2 m_e \beta^2\gamma^2\ov 1+ 2\gamma  m_e/m_w+m_e^2/m_w^2}  
\ee
Therefore we compute the energy lost by the ionizing particle in this region by integrating eq.(\ref{lostr})
\be
\int_{|t|_1}^{|t|_{max}} d|t| {|t|\ov 2 m_e} {d\sigma\ov dt}\rho_N = {2\pi\alpha^2\ov  m_e \beta^2}\big(\log({2 m_e Q_{max}\ov q_1^2})-\beta^2+{Q^2_{max}\ov 4\gamma^2 m_w^2} \big)\rho_N \label{lost2}
\ee
putting to zero the lower limit $|t|_1$ except than in the logaritmic divergent term.
 
 By summing the results eq(\ref{lost1}) and eq(\ref{lost2}) we finally get
 \be
 {dE\ov dx} =-{2\pi\alpha^2\ov  m_e \beta^2}\big(\log({2 m_e \beta^2\gamma^2Q_{max}\ov I^2})-2\beta^2+{Q^2_{max}\ov 4\gamma^2 m_w^2} \big)\rho_N
 \label{be}
 \ee     

 Remember that  $\rho_N={ \rho Z \ov A m_{a}}$ where $\rho$ is the mass density of the material, $Z$ and $A$ its atomic number and its atomic weight, and $m_{a}$ is the proton mass, neglecting the neutron-proton mass difference, or the atomic weight unit that is $1/12$ of the mass of $^{12}C$. In our computation we express $\rho_N/m_e$ in MeV/cm.
 
 Eq.(\ref{be}) is a differential equation, since $\beta,\gamma=\beta(x),\gamma(x)$ in the r.h.s are computed from $E(x)$: \\
$\gamma(x)={E(x)\ov m_w} ~~~~ \beta(x)=\sqrt{1-1/\gamma(x)^2}$.
 
 Eq.(\ref{be}) is the modern version of the so-called Bethe-Block formula, as it appears in the modern literature, up to two additional, less crucial terms, which we take from the literature.
 
 A term   ${2\pi\alpha^2\ov  m_e \beta^2}\delta$  representing a kind of screening effect whose expression for large energy is \cite{passage},\cite{Striganov}
 \be
 \delta=2 \log[\sqrt{4\pi\rho_N \alpha/m_e}/ I]+\log[\beta\gamma]-1/2 \label{delta}
 \ee
As it is said to be relevant for  $\beta\gamma>4$ (see Fig.32.1 of  \cite{passage} and Fig.1 of \cite{Striganov}), we use the expression (\ref{delta}) multiplying it by the pre-factor
$h(\beta)={\beta\gamma/4\ov 1+\beta\gamma/4})$.
 
 Further, a term $-{\alpha^3\ov 2 m_e }\Delta R$ representing the energy lost by photon radiation  \cite{Striganov}, with
 \be
 \Delta R=(\log[2\gamma]-{1\ov 3}\log[{2Q_{max}\ov m_e}])\log[{2Q_{max}\ov m_e}]^2
 \ee

 Summarizing, the formula that we have used in our computation is the same as eq.(15) of ref(\cite{Striganov}) (up to our pre-factor $h(\beta)$) namely\footnote{see also eq.(32.5) of \cite{passage} which however does not contain $\Delta R$ nor the term ${Q^2_{max}\ov 4}$.}
 \be
 {dE\ov dx} =-{4\pi\alpha^2\ov  m_e \beta^2}\big ({1\ov 2}\log({2 m_e \beta^2\gamma^2Q_{max}\ov I^2})-\beta^2+{Q^2_{max}\ov 8 \gamma^2 m_w^2} -h(\beta){\delta\ov 2}+{\alpha \beta^2\ov 4\pi }\Delta R \big) \rho_N
 \label{befin}
 \ee
 
 As for the value of the $I$, we follow the Landau-Lifschitz Quanum Mechanics book \cite{LLB},  and put
 $I= Z \times10 eV$, which is also consistent with Fig.32.5 of ref(\cite{passage}) and Fig.3 of ref(\cite{Striganov}) (actually from these figures, we read 12eV for Z(Si)=14, rather than 10eV, the result for $dE/dx$ being quite the same).
 
\vskip0.5cm

We can also evaluate the fluctuations from eq.(\ref{E}).

Since the integral in the r.h.s. of eq.(\ref{E}) is non singular for $\Delta\to 0$, we can approximate it by taking the relativistic formula for the cross-section with the lower limit put to $0$.
Therefore\footnote{We use here the relativistic eq.(\ref{sigma}) rather than the non relativistic computation of  \cite{fano}. Note that in our case $Q_{max}$ is much larger than the electron atomic energies  and therefore 
${<|\Sigma v_i|^2>_0\ov \beta^2} << 1$ in eq.(72) of \cite{fano}.}

\be
{d~ var[E](x)\ov dx}= \int_0^{|t|_{max}} d|t| ({t\ov 2m_e})^2 {\pi \alpha^2\ov m_e^2 p^2}{1\ov t^2}(4 m_e^2 p_0^2-|t|s+{t^2\ov 2})\rho_N 
=  {4\pi\alpha^2 \gamma^2\ov a}(1-{\beta^2\ov 2}+{2 m_e^2 \beta^4\gamma^2\ov 3 m_w^2 a^2})\rho_N   \label{varian}
\ee
where $a=1+{m_e^2\ov m_w^2}+2 {m_e\ov m_w}\gamma$.  Here, the Wino velocity  appearing in the r.h.s. is supposed to be a function of $x$: $\beta(x),\gamma(x)$ 
determined by the solution of eq.(\ref{befin}). Here we express $\rho_N$ in MeV$^2$/cm.

U.Fano \cite{fano} points out that the probability distribution is expected to be gaussian whenever the energy lost in the interval $\Delta(x)$ is much larger than the maximum loss in a single collision, and therefore in this case the Bethe-Block equation should be adequate, the uncertainty being estimated by the variance eq.(44). 
We find this to be our case.
Also, ref(\cite{Striganov}) points out in pag.3 that "straggling" (that is deviation from gaussian) is less important for heavy particles.

\vskip0.1cm

However, in the literature there is some debate whether, for thin silicon detectors, the fluctuations are gaussian distributed or 
there are long tails and it would be better to use the formula for the most probable energy loss  $\Delta E_{most}$ (MeV) in a width $\Delta x$ (cm) (see ref(\cite{passage}) eq.(32.11))
\be
\Delta E_{most}={2\pi\alpha^2\ov  m_e \beta^2}\rho_N \Delta x \big( \Log[{\gamma^2 4\pi\alpha^2\rho_N \Delta x\ov I^2}]+0.2-\beta^2-\delta \big) \label{most}
\ee

We have also used this formula to check the robustness of our results.

\end{document}